\documentclass[
aps,prd,
nofootinbib,
superscriptaddress,
showpacs,
tightenlines,
]{revtex4}
\usepackage{amsmath}
\usepackage{amssymb}
\usepackage{bm}
\usepackage{color,graphicx}
\usepackage{subfigure}
\usepackage{multirow}

\newcommand{\eq}[1]{Eq.(\ref{#1})}

\newcommand{\ba}{\begin{eqnarray}}
\newcommand{\ea}{\end{eqnarray}}

\newcommand\ddfrac[2]{\frac{\displaystyle #1}{\displaystyle #2}}

\def\be{\begin{eqnarray} }

\def\ee{\end{eqnarray}}

\def\bq{\begin{eqnarray}}
\def\eq{\end{eqnarray}}

\usepackage{color}
\usepackage{lineno,soul,xspace}
\definecolor{chromeyellow}{rgb}{1.0, 0.65, 0.0}

\begin{document}

\title{Pion nucleus Drell-Yan process and 
parton transverse momentum in the pion}

\author{Federico Alberto Ceccopieri} 
\email{federico.alberto.ceccopieri@cern.ch}
\affiliation{
IFPA, Universit\'e de Li\`ege, B4000, Li\`ege, Belgium}

\author{Aurore Courtoy} 
\email{aurore@fisica.unam.mx}
\affiliation{
Instituto de F\'isica, Universidad Nacional Aut\'onoma de M\'exico\\
Apartado Postal 20-364,  01000
Ciudad de M\'exico, Mexico. }

\author{Santiago Noguera}
\email{santiago.noguera@uv.es}
\affiliation{Departament de Fisica Te\`orica and IFIC, Universitat de Val\`encia- CSIC\\ 46100 Burjassot, Spain } 

\author{Sergio Scopetta}
\email{sergio.scopetta@pg.infn.it}
\affiliation{Dipartimento di Fisica e Geologia, Universit\`a degli Studi di Perugia and Istituto Nazionale di Fisica Nucleare, Sezione di Perugia\\ via A. Pascoli, I - 06123 Perugia, Italy.}

%
\pacs{...}

\begin{abstract}
We present a thorough analysis of unpolarized Drell-Yan (DY) pair production in 
pion-nucleus scattering. On the nucleus side, we use nuclear parton distributions along with 
parametrisations of the nucleon partonic transverse distribution available in the literature. Partonic longitudinal and transverse distributions of the pion are those obtained in a recent calculation in a Nambu-Jona Lasinio (NJL) framework,
with Pauli-Villars regularization. The scale of the NJL model is  determined
with a minimisation procedure comparing NLO predictions based on NJL evolved pion distributions to 
rapidity differential DY cross sections data. The resulting distributions
are then used to describe, up to next-to-leading logarithmic accuracy, 
the transverse momentum spectrum of dilepton 
pairs up to a transverse momentum of 2 GeV.
With no additional parameters, fair agreement is found
with available pion-nucleus data,
confirming the virtues of the NJL description of pion parton structure. 
We find sizable evolution effects on the shape of the distributions and on the generated average
transverse momentum of the dilepton pair.
 We furthermore discuss the possibility of gaining information about the behavior of the pion unpolarized transverse momentum dependent parton distribution from pion nucleus DY data.
 \end{abstract}

\maketitle
%
\baselineskip 3.0ex

\section{Introduction}

The non perturbative transverse structure of hadrons
has attracted recently much attention and the 
issue of extracting transverse momentum dependendent
parton distributions (TMDs) from data taken in different processes
in present and forthcoming high-luminosity facilities
represent{\bf s} an important goal of nowadays hadronic Physics.
In particular,
Drell-Yan (DY) pair production~\cite{Drell:1970wh}, discussed in this paper,
and semi-inclusive deep inelastic scattering are the main processes
under investigation~\cite{Diehl:2015uka}.

The cross section for DY pair production, differential in the transverse momentum of the pair, $q_T$,
is a particularly suitable observable for this kind of studies. 
In particular at small $q_T$, where the TMD formalism is formulated, 
fixed order calculation of this process show 
large logarithmic corrections due to an incomplete
cancellation of soft and collinear singularities between real and virtual
contributions and need to be resummed to all orders
to recover the predictivity of the theory~\cite{DDT,PP,Altarelli:1984pt}. 

The description of the $q_T$ DY spectrum in $pp$ collisions has reached a high degree
of sophistication~\cite{CSS}. On one side, theoretical improvements have increased the 
perturbative accuracy of the predictions~\cite{Bozzi:2008bb,Bozzi:2010xn,Catani:2015vma,Catani:2000vq,deFlorian:2000pr}.  On the other side, global fits 
of DY production at different energies have given access to the non perturbative 
proton transverse structure~\cite{BLNY,KN05}. Both aspects have received increasing attention due to the formalisation of new and old concepts in the TMD language~\cite{AR,Bacchetta:2017gcc,Scimemi,Su:2014wpa}.  
While  there are differences between the language used in the modern and the older TMD approaches, 
 physical results should not depend on 
 it. A detailed comparison of the formalisms
can be found in Refs.~\cite{Prokudin:2015ysa,Ceccopieri:2014qha}.

At high energy colliders, this improved knowledge aims to an increasingly better 
description of electroweak bosons production, with the Higgs $q_T$ spectrum 
being the highlighted case. Measurements of $q_T$ spectrum of the DY process, at lower centre of mass energies, 
are instead more sensitive to the  hadronic non perturbative transverse structure.     

DY pair production in pion-nucleus scattering
is a unique probe of pion parton distribution functions
(PDFs) and, as such represents a source of information on 
the pion parton structure. 
In particular for the $q_T$ spectrum this was realized long time ago
by the authors of Ref.~\cite{Chiappetta:1981vv}. More recently, phenomenological analyses 
have appeared~\cite{Pasquini:2014ppa}. A fit to the $q_T$ spectrum 
of DY pairs produced in pion-nucleus collisions has been  recently presented
in Ref.~\cite{Wang:2017zym}.

Pion TMDs, which could be extracted in principle
in a next generation of 
pion-nucleus DY experiments \cite{COMPASSII-proposal},
have received recently considerable theoretical interest,
\cite{Engelhardt:2015xja,Musch:2011er,
Lu:2004hu,Gamberg:2009uk,Lu:2012hh,Pasquini:2014ppa,ns,Bacchetta:2017vzh}.
In this paper we study the DY unpolarized
pair production in pion-nucleus scattering,
to next-to-leading logarithmic (NLL) perturbative accuracy, 
up to a transverse momentum of the produced lepton pair
of 2 GeV.
As non perturbative inputs, we use, for the bound nucleons,
a longitudinal structure which takes into account
nuclear effects and, for the
transverse structure, a well established parameterization
obtained through 
a phenomenological fit to proton-proton DY data (called, from now on,
KN05 prescription) \cite{KN05}.
For the pion, we use TMDs obtained in
a recent calculation \cite{ns}, within a Nambu-Jona Lasinio (NJL) framework
\cite{Klevansky:1992qe},
with Pauli-Villars regularization.
The corresponding RGE scale of the model is determined in a novel way by comparing the DY unpolarized cross section, integrated over $q_T$, described by evolved pion PDFs evaluated in the NJL model to the data.

The aim of the present paper is  to study the performances of the NJL model, widely used to describe
the non-perturbative meson structure, against DY differential cross 
section data for the first time.
 We also analyze to what extent this process 
can be used to obtain information
on the pion transverse structure in momentum space, as
it happens for the proton in the corresponding process.

The paper is structured as follows. In the next section, we present the set-up of the 
calculation and introduce the ingredients used to describe
the proton and pion structure.
In the third section, we discuss the results of the calculation of DY cross sections
in the kinematics of presently available data for pion-tungsten scattering. Eventually, we draw our conclusions in the last section.

\section{Setting-up the calculation}

\subsection{Drell-Yan cross section}
\noindent
In the following we will be interested in the process of the type
\begin{equation}
\label{processDY}
h_1(p_1) \; h_2(p_2)\to \gamma^*(q)+X,
\end{equation}
in which a virtual photon is produced with large 
invariant mass $Q^2$ and transverse momentum $q_T$ in the 
collisions of two hadrons at a centre-of-mass energy $s = (p_1 + p_2)^2$, with $p_{1,2}$ the four momentum of hadrons $h_{1,2}$, respectively.
When $q_T^2$ becomes small compared to $Q^2$, large logarithmic 
corrections of the form of $\alpha_s^n \log^m(Q^2/q_T^2)$ with $0 \leq m \leq 2n-1$
appear in fixed order 
results, being $n$ the order of the perturbative calculation. 
These large logarithmic corrections can be resummed to all
orders by using the Collins-Soper-Sterman (CSS) formalism~\cite{CSS}.
In this limit, of interest for the present analysis and neglecting finite corrections in the $q_T\sim Q$ region, the 
cross-section can be written as 
\begin{eqnarray}
\label{CSS}
\frac{d\sigma}{dq_T^2 d\tau dy}&=& \sum_{a,b} \sigma_{q\bar{q}}^{(LO)}  
\int_0^\infty db \frac{b}{2} J_0(b \, q_T)\,S_q(Q,b)\, S_{NP}^{h_1 h_2}(b) \cdot \nonumber\\
&& \cdot \Big[ \left(f_{a/h_1}\otimes C_{qa}\right)\left(x_1,\frac{b_0^2}{b^2}\right) \,
\left(f_{b/h_2}\otimes C_{\bar{q}b}\right)\left(x_2,\frac{b_0^2}{b^2}\right)
+ q \leftrightarrow  \bar q \Big]\,.
\end{eqnarray}
where $b_0=2 e^{-\gamma_e}$, the symbol $\otimes$ stands for convolution 
and $\sigma_{q\bar{q}}^{(LO)}$ is the leading-order 
total partonic cross section for producing a lepton pair, 
$\sigma(q \bar q \rightarrow l^+ l^-)$,
and it is given by 
\bq
\sigma_{q\bar{q}}^{(LO)}   = {4 \pi \alpha_{em}^2 \over 9 Q^2} e_q^2\,.
\eq
In Eq.~(\ref{CSS}), the $a,b$ indices run on quark and gluons, 
$J_0(b \, q_T)$ is the Bessel function of first kind and $f_{i/h}$ corresponds to 
the distribution of a parton $i$ in a hadron $h$. 
The cross section in Eq.~(\ref{CSS}) is differential in $\tau=Q^2/s$ and $y$, the rapidity of the DY pair. 
Momentum fractions appearing in parton distribution functions 
can be expressed in terms of these variables as
\bq
x_{1(2)}
= \sqrt{\tau} e^{\pm y},  \,\,\,\,\,\,\,\,  y=\frac{1}{2} \ln \frac{x_1}{x_2}\,.
\label{kine}
\eq
Cross sections differential in $x_F=x_1-x_2= 2 q_\parallel / \sqrt{s}$, the longitudinal momentum of the pair in the hadronic centre of mass system, 
can be obtained from those differential in rapidity $y$ by a suitable transformation.
By defining $A=\sqrt{x_F^2 + 4 \tau}$ one gets
\bq
x_1= {x_F \over 2} + {A \over 2}, \; \; \; \; \; \; \; \; \; \; \; \;
x_2= -{x_F \over 2} + {A \over 2}, \;\;\; \; \; \;\; \; \; \; \; \; dy     = dx_F / A \,.
\label{kine2}
\eq
Momentum conservation further imposes that $|x_F|<1-\tau$.
The large logarithmic corrections are conventiently exponentiated 
in $b$-space in the Sudakov perturbative form factor
\begin{equation}
\label{sudakov}
S_q(Q,b)=\exp \left\{ -\int_{b_0^2/b^2}^{Q^2} \frac{dq^2}{q^2} 
\left[ A(\alpha_s(q^2)) \;\ln \frac{Q^2}{q^2} + B(\alpha_s(q^2)) \right] \right\}\, .
\end{equation}
The functions $C_{ab}$ in Eq.~(\ref{CSS}) and 
$A$, $B$ in Eqs.~(\ref{sudakov}) have perturbative
expansions in $\alpha_s$,
\begin{eqnarray}
\label{aexp}
&&A(\alpha_s) = \sum_{n=1}^\infty \left( \frac{\alpha_s}{2\pi} \right)^n A^{(n)}, \;\;
B(\alpha_s) = \sum_{n=1}^\infty \left( \frac{\alpha_s}{2\pi} \right)^n B^{(n)}
\;\;, \\
\label{cexp}
&&C_{ab}(\alpha_s,z) = \delta_{ab} \,\delta(1-z) + 
\sum_{n=1}^\infty \left( \frac{\alpha_s}{2\pi} \right)^n C_{ab}^{(n)}(z) \;\;.
\end{eqnarray}
At present, the perturbative Sudakov form factor can be evaluated
at next-to-next-to-leading logarithmic (NNLL) accuracy~\cite{deFlorian:2000pr}. 
In the $q\bar{q}$ annihilation channel pertinent to Drell-Yan production,
the evaluation of the Sudakov form factor at next-to-leading logarithmic (NLL) accuracy,
the one reached in the present analysis,
involves the coefficients
\begin{equation}
A^{(1)}=2 C_F \;\;\;\; B^{(1)}=-3 C_F \,,
\end{equation}
which are the coefficient of the singular $(1-z)^{-1}$ and $\delta(1-z)$ terms of the one-loop 
splitting function $P_{qq}^{(0)}(z)$ and 
\begin{equation}
A^{(2)}= K A^{(1)}    \;\;\;\; K=C_A \left( \frac{67}{18}- \frac{\pi^2}{6} \right) - n_f T_R \frac{10}{9} \,,
\end{equation}
which is the coefficient of the singular term of the two-loop 
splitting function $P_{qq}^{(1)}(z)$ in the $z\rightarrow 1$ limit~\cite{KT}.
The general expression for $C_{ab}^{(1)}$ are given by~\cite{DS,deFlorian:2000pr}
\begin{eqnarray}
\label{coeff}
C^{(1)}_{qa} (z)=C^{(1)}_{\bar{q}b} (z)&=& \delta_{qa} C_F\, (1-z) + \delta_{qa}\, \delta(1-z) C_F 
\left( -4 + \frac{\pi^2}{2} \right)\,,\nonumber\\
C^{(1)}_{qg} (z)=C^{(1)}_{\bar{q}g} (z)&=& 2 T_R\,  z (1-z)\,.
\end{eqnarray}
Color factors in the previous equations are given by $C_A=3$, $C_F=4/3$, $T_R=1/2$ with $n_f$ being the number of active flavours.
Together with the use of NLO pdfs, this guarantees the evaluation of the cross section 
at small $q_T$ at NLL accuracy.  
The last ingredient in Eq.~(\ref{CSS}) is the non perturbative
form factor, $S_{NP}^{h_1 h_2}(b)$, which encodes the transverse structure
of both the colliding hadrons. The latter is either fixed by comparison with data or
parametrized with the help of hadronic 
models, as we shall do in this paper.

\subsection{Proton structure}

Predictions for the transverse momentum spectrum of DY pairs produced in 
pion-proton collisions do rely on the knowledge of the proton NP form factor.
The latter is extracted from the transverse momentum spectrum of DY pairs produced in proton-proton ($pp$) and proton-nucleus ($pA$) collisions. 
Quite recent analyses~\cite{Scimemi,Bacchetta:2017gcc} have appeared which address such an extraction.
Since our aim here is   to establish the possibility of studying the pion transverse
non perturbative structure in pion-nucleus DY experiments, we here intend to minimize the uncertainity coming from the proton structure part of the calculation.
 We use the
well known and widely accepted results of Konychev and Nadolsky (KN05)~\cite{KN05} 
obtained within the CSS formalism~\cite{CSS} where $S_{NP}^{pp}(b)$ is extracted from global fit to $Z$-boson and low mass DY data, updating the results presented in Ref.~\cite{BLNY}.
\begin{figure}[t]
\begin{center}
\includegraphics[scale=0.6]{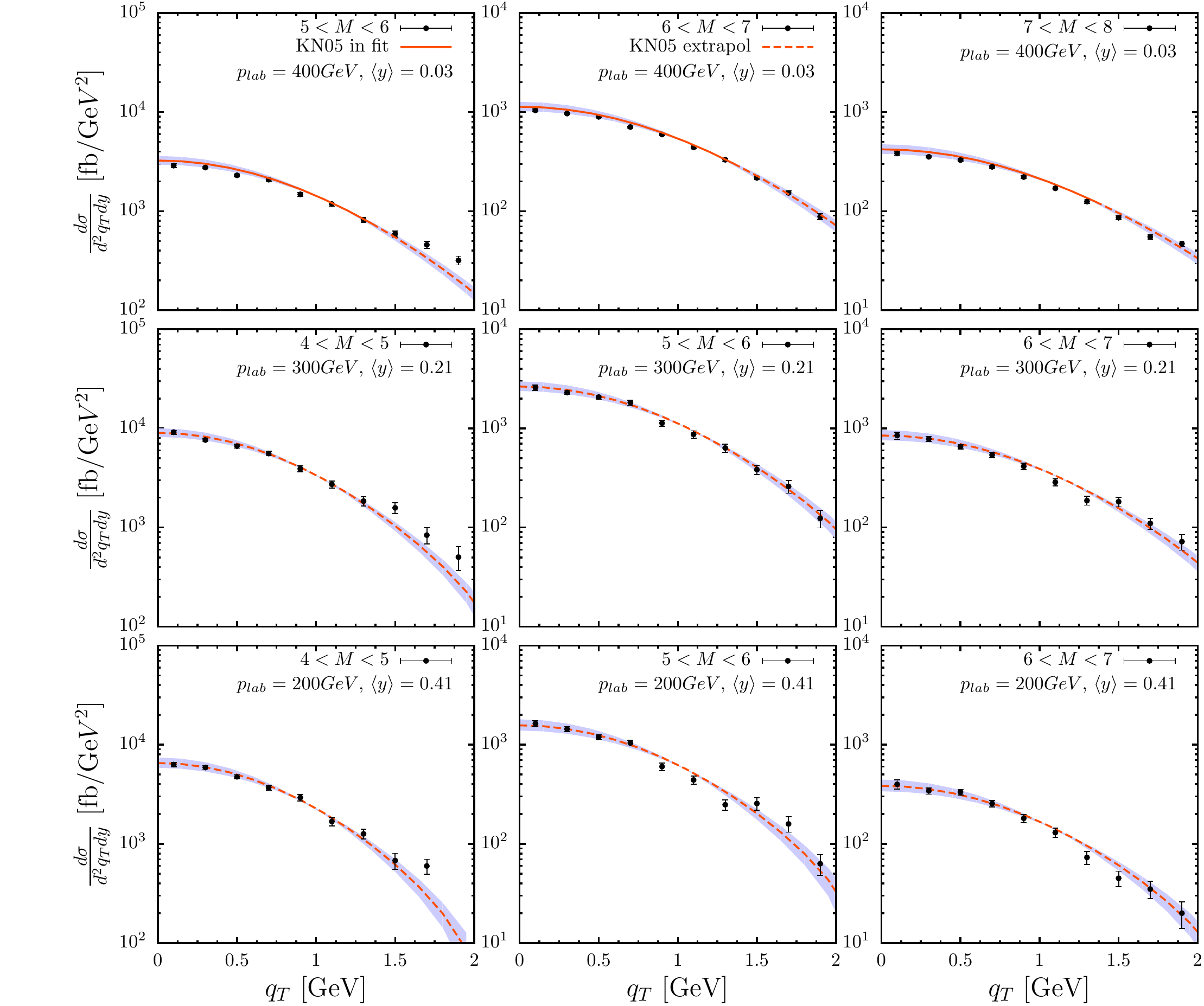}
\caption{Theoretical predictions obtained with the KN05 model~\cite{KN05} compared 
to DY transverse momentum spectra in $pA$ collisions~\cite{Ito} in bins of the invariant mass of the pair, $M$, expressed in GeV,
for different incident beam energies and DY pairs rapidities.
Solid lines indicate predictions in the phase space region included in the KN05 fit
whereas dashed ones indicate predictions in 
an extrapolation regime. The error band corresponds to the $a_i's$ error propagation.}
\label{Fig:1}
\end{center}
\end{figure}
The latter is parametrised as
\begin{equation}
S_{NP}^{pp}(b)=\exp\{-[a_1 + a_2 \ln (M/(3.2 \,\mbox{GeV})) + a_3 \ln(100 x_1 x_2)] b^2\}\,.
\label{fnp}
\end{equation}
The $a_i$ parameters appearing in Eq.~(\ref{fnp}) are determined by a minimisation procedure 
against data and are given by~\cite{KN05}
\begin{equation}
a_1=0.201 \pm 0.011, \;\; a_2=0.184 \pm 0.018, \;\; a_3=-0.026 \pm 0.007 \,.
\end{equation}
The fit is fully specified once a prescription for the treatment of the 
non perturbative, large-$b$, region both in the Sudakov form factor, 
Eq.~(\ref{sudakov}), and the parton distributions
is given. The authors of Ref.~\cite{KN05} adopt the so-called
$b_\star$-prescription, substituting $b$ with
\begin{equation}
b_\star(b,b_{max})=\frac{b}{\sqrt{1+\Big(\frac{b}{b_{max}}\Big)^2}} \,,
\end{equation}
and setting $b_{max}=1.5 \, \mbox{GeV}^{-1}$ in the perturbative form factor.
In principle, the same setting should be used in PDFs, which are evaluated 
at the factorisation scale $\mu_F=b_0/b_*$. 
However this choice for $b_{max}$ may imply a call to a specific
PDFs parameterization below their lowest available scale, $Q_{in}$.
Since in Ref.~\cite{KN05} cross sections are evaluated with the NLO CTEQ6M PDFs~\cite{CTEQ6}, 
whose lowest $Q$ accessible is $Q_{in}=1.3$ GeV, the $b_\star$-prescription entering PDFs calls 
is used with $b_{max}=b_0/Q_{in} \simeq 0.86 \, \mbox{GeV}^{-1}$ which always guarantees $\mu_F>Q_{in}$.
It is important to remark that 
the non perturbative form factor is determined not only by fitting the parameters 
of the chosen functional form, but also by the specific regularisation 
prescription and its associated parameters adopted to deal with the infrared region.   
In general all these ingredients have been found to be highly correlated.

In order to present a benchmark of our code and to gauge how theory 
performs in extrapolation regions,  
we compare predictions from KN05 to the $pA$ data of Ref.~\cite{Ito}.
An additional $\pm 25 \%$ normalisation error is assigned to the data~\cite{Ito}.
In the original KN05 analysis, only the data at $p_{lab}=400$ GeV, 
$q_T<1.4$ GeV, $5<M/\mbox{GeV}<9$ were included in the fit. In such a restricted region indeed the theory
(solid lines) performs well offering a good benchmark of our code, as shown in 
the first row of Fig.~\ref{Fig:1}. 
Since the $\pi W$ data to be analyzed in the following are at $p_{lab}$=252 GeV, 
it is important to check how well 
the theory performs in extrapolation regions at lower $\sqrt{s}$ and higher DY rapidity.
Therefore we present in the second and third rows of Fig.~\ref{Fig:1} 
the KN05 benchmark (dashed lines) versus data~\cite{Ito} at $p_{lab}$= 200 and 300 GeV, 
which were not included in the KN05 fit.
By using Eq.~(\ref{kine}) and Eq.~(\ref{kine2}) and assuming the invariant mass values indicated on the plots, 
the rapidity coverage of these data can be converted to the range $0<x_F<0.3$.
In both cases we find good agreement between data and theory up to $q_T \sim 2$ GeV
giving us confidence that the KN05 model can be successfully used in this {\bf ($x_F,q_T$)} range
at the $\sqrt{s}$ of interest in this analysis.

\subsection{Pion structure}

A calculation of pion TMDs in a NJL framework,
with Pauli-Villars regularisation,
has been recently presented
in Ref. \cite{ns}.
Model calculations of meson partonic structure within
this approach have a long story of successful predictions
\cite{Davidson:2001cc,Theussl:2002xp,
RuizArriola:2002bp,Noguera:2011fv,Weigel:1999pc,Broniowski:2017gfp}.
Collinear parton distributions 
obtained within a model have to be associated to a low 
momentum scale $Q_0^2$ and, 
in order to be used to predict measured quantities,
have to be evolved to higher momentum scales according to
perturbative QCD (pQCD).

In Ref. \cite{ns} the unpolarized NJL parton TMD has been obtained.
Among its good properties, we stress that, upon integration over the intrinsic quark
transverse momentum $\bm{k_T}$,
the pion PDF $q(x)$ is properly recovered with correct
normalisation and the momentum sum rule is exactly satisfied. This is due to the fact that NJL is
a field theoretical scheme and the correct
support of the PDF, $0\leq x\leq1$, is not imposed
but arises naturally. 
In particular, the momentum sum rule reads
$\int dx~x~q\left(  x\right)  =0.5$, {\it i.e.}
the fraction of momentum carried by each quark is one
half of the total momentum, since at the scale of the model
only valence quarks are present. 
The dependence on $\bm{k_T}$ of the TMD obtained in Ref. \cite{ns}
is very important for the present study.
It is worth stressing that, in this approach, the  $\bm{k_T}$ dependence is automatically generated by the NJL dynamics and it is not imposed by using any educated
guess. This is an important feature of the results of Ref. \cite{ns},
not found in other approaches \cite{Frederico:2009fk,Pasquini:2014ppa}.
In this paper we will use the pion TMD obtained
in Ref. \cite{ns} in the chiral limit, which is
an excellent
approximation to the NJL full result; this is very convenient
in the present calculation, since,
at the low but undetermined scale $Q_0^2$ associated to the model, the pion TMD 
can be written in a factorised form 
\bq
f^{q/\pi}(x_{\pi},\bm{k_T},Q_0^2)= q(x_{\pi},Q_0^2) T(\bm{k_T})~,
\label{chiral}
\eq
where one has (in $\pi^-$, of interest here):
\begin{equation}
\label{ic}
q(x_{\pi},Q_0^2)=d_v(x_{\pi},Q_0^2)=\bar{u}(x_{\pi},Q_0^2)=1\,.
\end{equation}
The function $T$ is given by
\begin{equation}
\label{Tfunction}
T(\bm{k_T})= 
\frac{3}{4 \pi^3} {\left( m \over f_\pi  \right )^2}~
\sum_{i=0,2} \frac{c_i}{k_T^2+m_i^2}\,,
\end{equation}
which, due to a proper combination of the $c_i$ \cite{ns}{\bf ,}
behave{\bf s} as $k_T^{-6}$ for asymptotic values of $k_T=|\bm{k_T}|$ and 
satisfies the normalisation
\bq
\label{norm}
\int d^2 \bm{k_T} \, T(\bm{k_T})= 1\,.
\eq
Since the distribution Eq. (\ref{Tfunction}) depends only upon $k_T^2$,
its Fourier transform can be cast in the form 
\begin{eqnarray}
S_{NP}^{\pi}(b)&=& \frac{3}{2 \pi^2} {\left( m \over f_\pi  \right )^2}~
\sum_{i=0,2} \int \, dk_T \, k_T \, J_0(b k_T)  
\frac{c_i}{k_T^2+m_i^2}  \nonumber\\
                &=& \frac{3}{2 \pi^2} {\left( m \over f_\pi  \right )^2}~
\sum_{i=0,2} c_i K_0 ( m_i \, b)\,,
\label{bsp}  
\end{eqnarray}
where $K_0$ is the modified Bessel function of the second kind. 
The parameters used in Eq.~(\ref{bsp}) are given in Ref.~\cite{ns} and read 
\begin{eqnarray}
&
m_0^2=m^2=(0.238\;\mbox{GeV})^2, \quad m_1^2=m^2+
\Lambda^2, \quad m_2^2=m^2+2\Lambda^2 
&\nonumber\\
&
\Lambda=0.860 \;\mbox{GeV},\quad
c_0=1, \quad c_1=-2, \quad c_2=1, \quad
f_\pi = 0.0924 \; \mbox{GeV}\,.
&\nonumber
\end{eqnarray}
As noted above, the NJL pion model  corresponds to a low
hadronic scale $Q_0^2$. Such a low scale has been determined previously 
by directly comparing the second moment of the pion PDF evaluated in NJL model with the results from the analysis of Ref.~\cite{Sutton:1991ay}. The procedure gives 
a value of $Q_0^2=0.18$ GeV$^2$ at NLO\footnote{Other schemes give higher values for the hadronic scale, {\it i.e.} up to $\sim 1$ GeV$^2$~\cite{Noguera:2005cc}.}~\cite{Courtoy:2008nf,CourtoyThesis}. 
%
\begin{figure}[t]
\begin{center}
\includegraphics[scale=0.6]{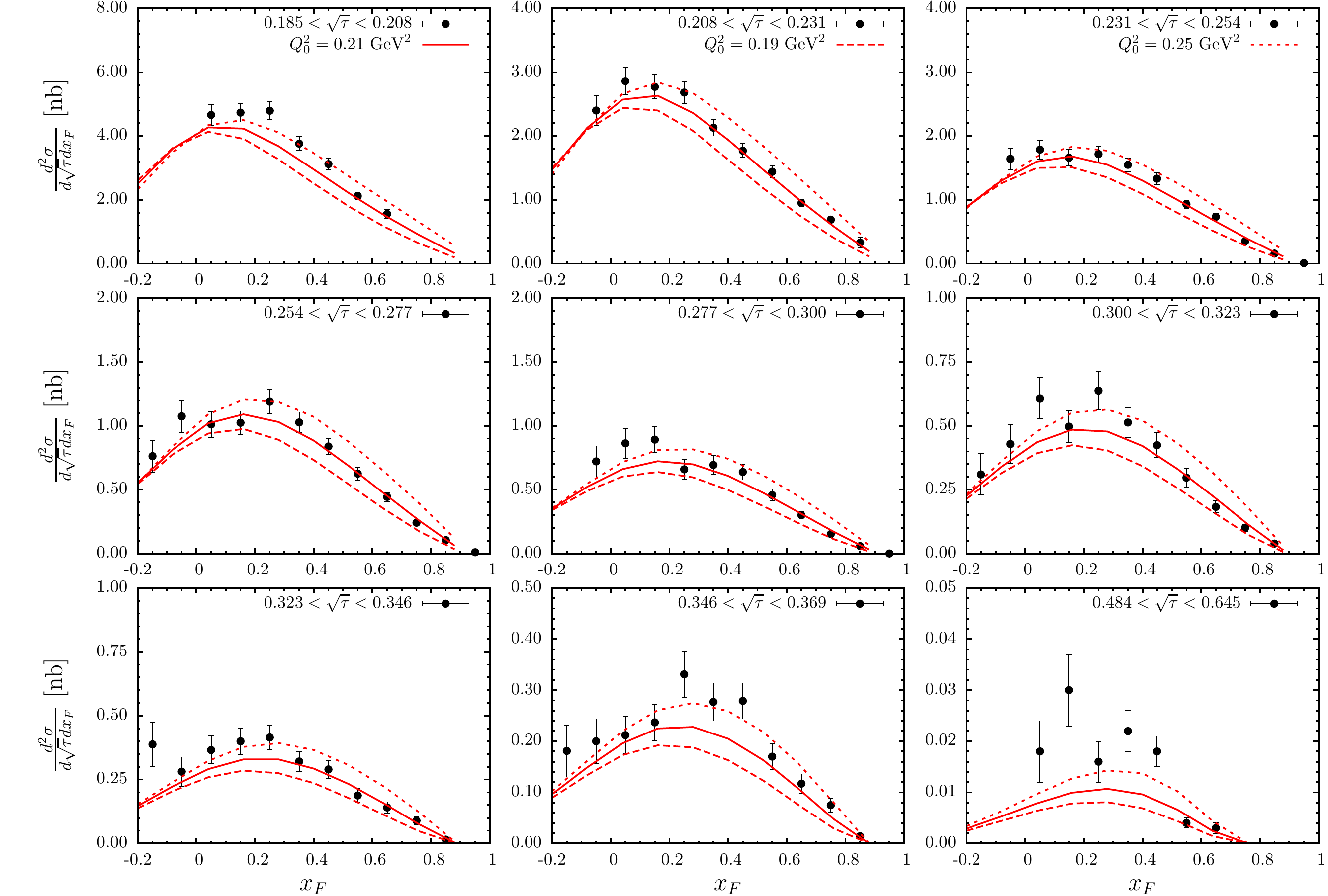}
\caption{Drell-Yan pairs production in $\pi^- W$ collisions. 
Next-to-leading order cross sections obtained by using evolved NJL pion PDFs 
for three values of $Q_0^2$ are compared to data of Ref.~\cite{conway}.}
\label{Fig:2}
\end{center}
\end{figure}
%
In the present paper we use  a different strategy : 
we consider $Q_0^2$ a free parameter of the NJL model which is then fixed with a minimisation procedure, outlined in the following, of the theoretical 
$\pi^- W$ DY cross sections, differential in $\sqrt{\tau}$ and $x_F$,
 against 
the corresponding experimental ones ~\cite{conway}.
Theoretical cross sections are calculated according to 
\begin{equation}
\frac{d^2 \sigma}{dQ^2 dx_F}=\frac{4\pi \alpha_{em}^2}{9 Q^2 s}
\sum_{ij} e_i^2 \int_{x_1}^{1} dt_1 \int_{x_2}^{1} dt_2
\frac{d^2 \hat{\sigma}^{ij}}{dQ^2 dx_F}
f_{i/\pi}(t_1,Q^2) f_{j/p}(t_2,Q^2)\,, 
\label{dsdxf}
\end{equation}
where the partonic cross sections $d\hat{\sigma}^{ij}$ are calculated at NLO accuracy 
by using the results of Ref.~\cite{Sutton:1991ay}. 
An additional correction takes into account 
the correct number of gluon polarisations in the $\overline{\mbox{MS}}$ in dimensional regularisation~\cite{Aicher:2010cb}.
The NJL pion PDFs are evolved to NLO accuracy in the Variable Flavor Number Scheme,
with the initial condition given in Eq.~(\ref{ic}), with the help of the 
QCDNUM~\cite{QCDNUM} evolution code. The QCD parameters 
are those of the NLO CTEQ6M parameterisation~\cite{CTEQ6}. 
In particular we set the NLO running coupling 
to $\alpha_s^{(n_f=5)}(M_Z)=0.118$ at the $Z$-boson mass, $M_Z$. 
Since the data we are comparing to are obtained on a tungsten target, 
we take into account nuclear effects by using nuclear PDFs 
of Ref.~\cite{nCTEQ}.
We have carried out a $\chi^2$ study to establish the hadronic scale of the model that describes the best the data at NLO in pQCD. Two cases have been considered: an evaluation of the $\chi^2$ for the full range of $x_F$ and another one with a cut $x_F<0.4$, since the NJL model is expected to better reproduce the pion valence distributions, expected to populate the range of large and positive $x_F$ . 
The scales thus determined are
\begin{equation}
Q_{0,\,\mbox{\tiny no cut}}^2=0.212_{-0.012}^{+0.011} \; \mbox{GeV}^2, \quad  \quad 
Q_{0,\,\mbox{\tiny cut}}^2=0.209_{-0.009}^{+0.008} \; \mbox{GeV}^2,
\end{equation}
and correspond to a chisquare value of $\chi^2/$d.o.f.$=2.1$ and $1.9$, respectively. The quoted errors correspond to a variation of one unit in $\chi^2$, {\it i.e.} 1-$\sigma$. Those results are compatible with each other.
We will therefore refer to $Q_0^2=0.21$ GeV$^2$,  as the scale associated to the pion NJL model.
The other two curves in Fig.~\ref{Fig:2}, corresponding to $Q_0^2=0.19$ GeV$^2$ 
and $Q_0^2=0.25$ GeV$^2$ respectively, are  added, 
in order to show the sensitivity to this particular choice 
of infrared $Q_0^2$. 
It is worth noticing that the results show an acceptable agreement, both in shape and in normalisation. 
More in detail, a tendency of the theory 
to undershoot the data is identified in the range of small 
$x_F$ ($-0.2<x_F<0.2$). 
This deficiency is not unexpected since, in the mentioned kinematic region, 
the dominant contribution
to the cross sections involves sea quarks and gluons which are
absent at  $Q_0^2$ and are  radiatively generated by 
QCD evolution. 
This is a typical drawback of models which contain only valence contributions
at the hadronic scale. At this point we would like to mention that the 
theoretical description of the $x_F$-spectra at large $x_F$ 
and the determination of pion parton distributions
can be further 
improved employing resummation techniques presented in Ref.~\cite{Aicher:2010cb,Aicher:2011ai,Westmark:2017uig}. It is worth noticing that, as shown in those papers, threshold NLL resummation of the Wilson coefficients leads to larger cross sections at large $x$ with respect to NLO ones. This{\bf ,} in turn{\bf ,} implies softer pion PDFs at large $x$. In the present context{\bf ,} this fact would imply a scale $Q_0^2$ for the NJL model  lower than the one already determined by using NLO Wilson coefficients in Eq. (\ref{dsdxf}).

\begin{figure}
\begin{center}
\includegraphics[scale=0.45]
{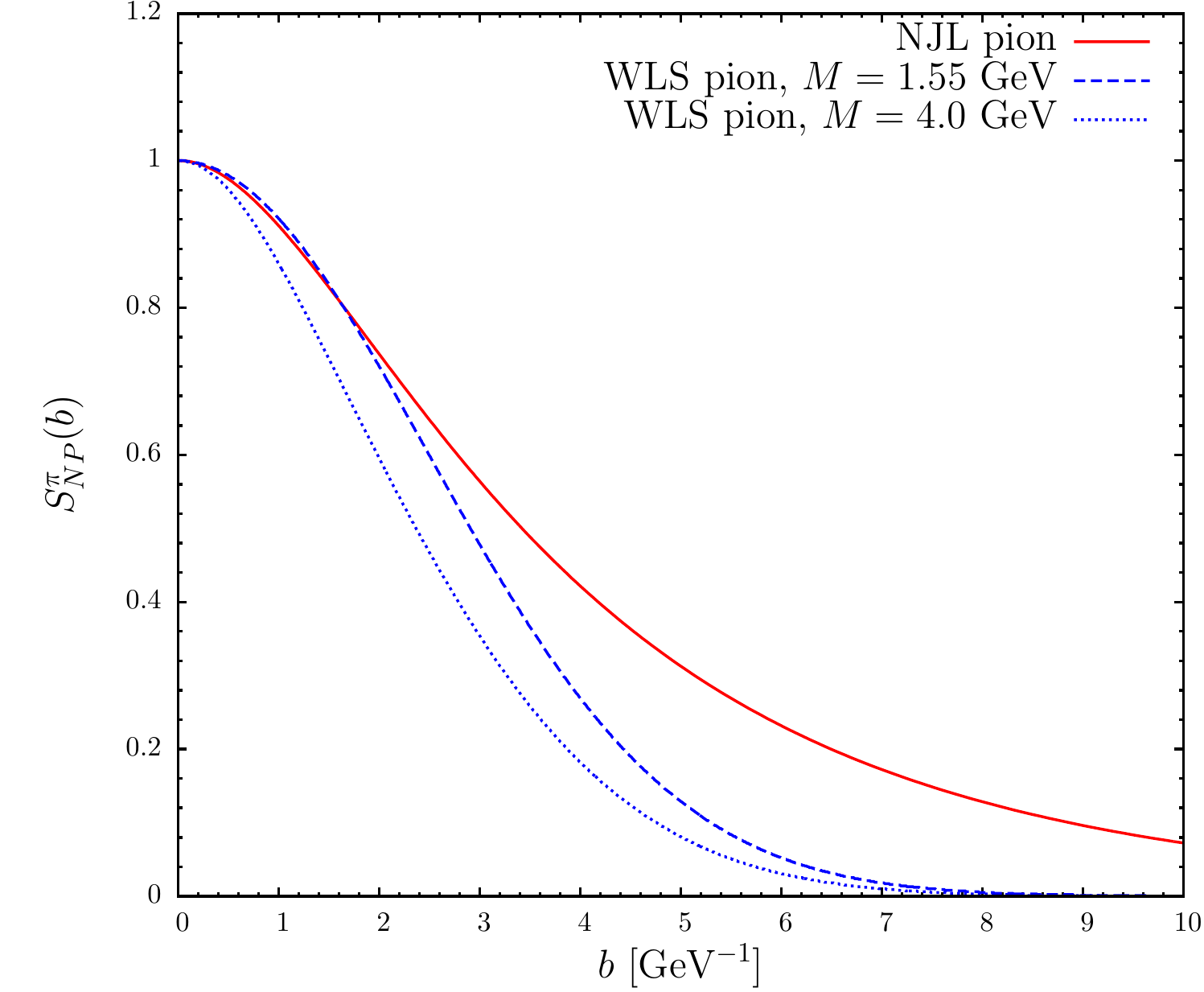}
\includegraphics[scale=0.45]{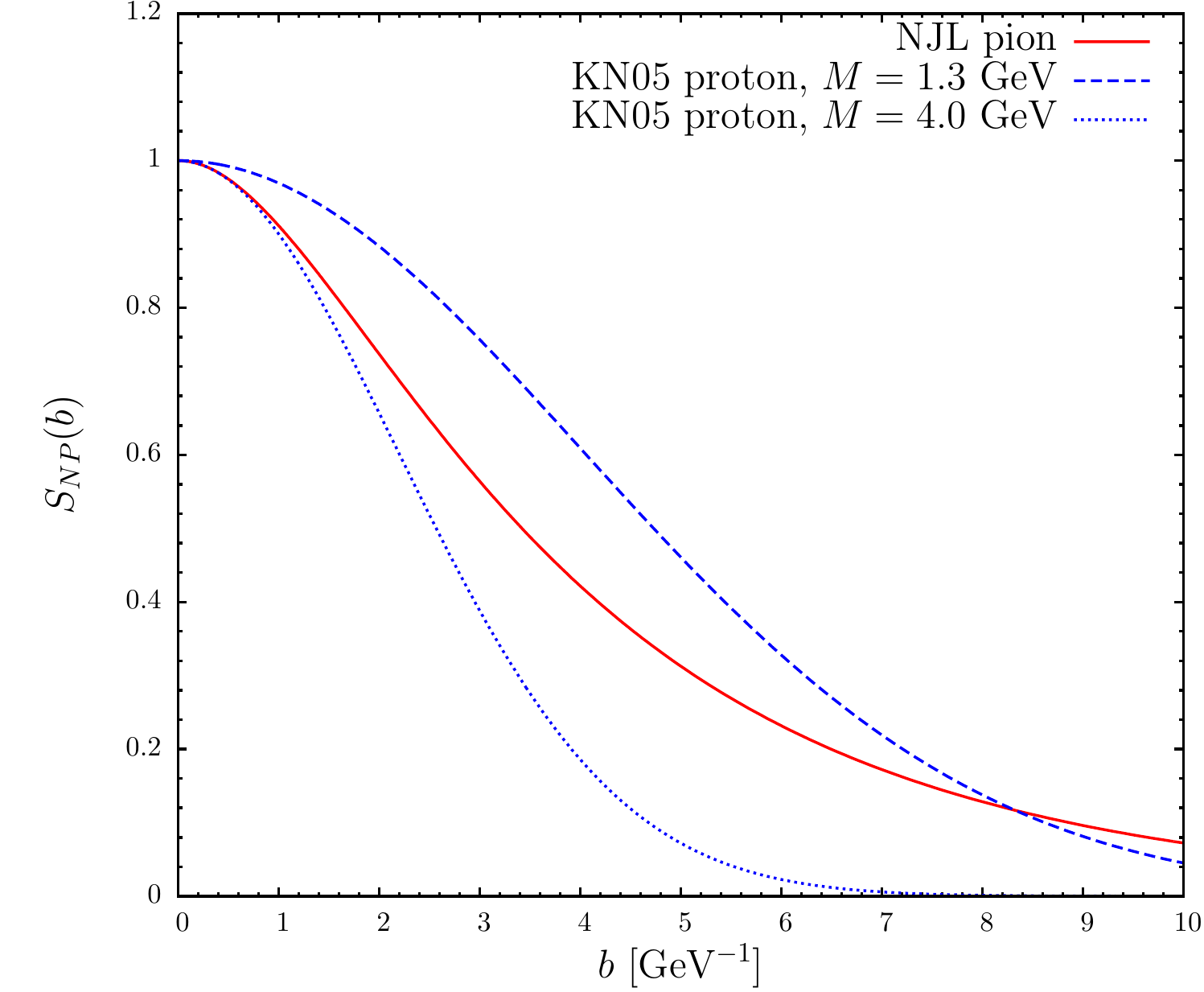}
\caption{Transverse profile in $b$-space for the NJL pion, eq.~(\ref{bsp}),
compared, in the left panel, to WLS pion~\cite{Wang:2017zym} and, in the 
right panel, to the KN05 proton, $\sqrt{S_{NP}^{pp}(b)}$, both evaluated for different values of the scale $M$.}
\label{Fig:3}
\end{center}
\end{figure}

\section{Predictions for $\pi W$ collisions data}
Predictions for the $\pi W$ Drell-Yan cross sections are obtained 
once appropriate modifications are implemented in Eq.~(\ref{CSS}). 
Evolved NJL pion parton distributions replace proton PDFs for hadron 1. 
Moreover the non-perturbative form factor $S_{NP}^{h_1 h_2}(b)$ depends on the particle species
initiating the reaction. Therefore in $\pi W$ collisions the latter is 
written as follows:  
\be
S_{NP}^{\pi W}(b)=S_{NP}^{\pi}(b) \, \sqrt{S_{NP}^{pp}(b)} \,,
\eq
where $S_{NP}^{\pi}(b)$ is given in Eq.~(\ref{bsp}) and the square root on $S_{NP}^{pp}(b)$, given 
in Eq.~(\ref{fnp}), takes into account that now only one proton is involved in the process.
It is instructive to directly compare 
the proton and pion non perturbative transverse distributions 
used in the calculation. 
It is important to remark that the NJL pion transverse distribution
in Eq.~(\ref{bsp}) differs from the corresponding proton factor in Eq.~(\ref{fnp}) in that 
it does not contain any explicit dependence neither on hard scale $M$ nor on parton fractional momenta.
Such a comparison is meaningful at the typical scale for which the transverse form factors
and the longitudinal momentum part
factorize. For the pion case this happens at the scale 
$Q_0^2$ determined in the previous section.
For the proton TMD such a scale is ambiguously defined and, according to KN05 analysis, ranges 
between $Q_{in}^2$ and $(b_0/b_{max}^{KN05})^2$. 
Therefore we choose $M=Q_{in}=1.3$ GeV in Eq.~(\ref{fnp}) and fix 
the product $x_1 x_2=M^2/s$, see Eq.~(\ref{kine}), exploiting the $\pi^- W$ kinematics with $s$ calculated according 
to a beam energy of $p_{lab}=252$ GeV. The comparison is presented in Fig.~\ref{Fig:3}. 
All the distributions reduce to unity in the $b \rightarrow 0$
limit, since they are all normalised to unity in transverse momentum space. 
In the left panel of Fig.~\ref{Fig:3} we compare the NJL transverse distribution
to the pion parametrisation of Ref.~\cite{Wang:2017zym} (called hereafter WLS) obtained from a fit 
of the same cross section 
data used in the present paper.  One may notice that, for this model, the width of the distribution
is smaller with respect to the NJL one, implying a larger average transverse momentum.
In the right panel of Fig.~\ref{Fig:3} one may notice that the NJL pion transverse distribution develops a larger tail with respect to the gaussian drop of the proton distributions.
Moreover the $b$-space width of the KN05 proton with $M=1.3$ GeV is larger with respect to the pion one. 
When transformed back in $k_T$ space, this implies that the intrinsic transverse momentum
in the pion is larger than the one in the proton, in agreement with the   
general expectations, since the pion is a much smaller system with respect to the proton.
It is worth mentioning that both the KN05 and WLS non perturbative form factors have an explicit, althought slightly different, dependence upon the hard scale $M$, in both cases set equal to the invariant mass of the dilepton pair.
Therefore we plot in each panels, as a representative case, the curves corresponding to both form factors evaluated with the scale set to $M=4$ GeV. Comparing the latter curves to the ones with $M \sim 1$ GeV, we conclude that the $M$-dependence generates a sizable non perturbative 
evolution  of the form factor which is more pronounced for KN05 proton model than for the WLS pion model. 

\begin{figure}[t]
\begin{center}
\includegraphics[scale=0.45]{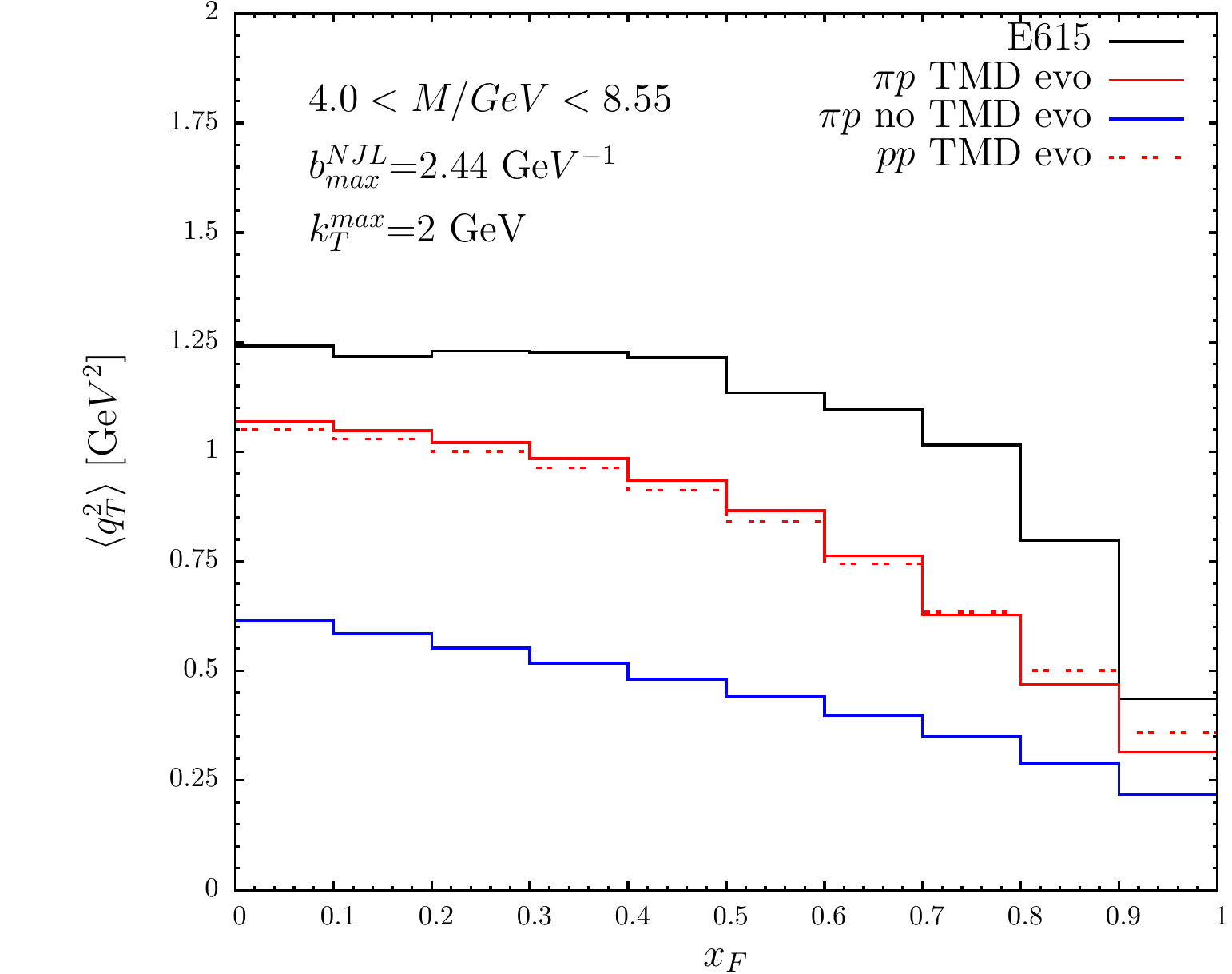}
\includegraphics[scale=0.45]{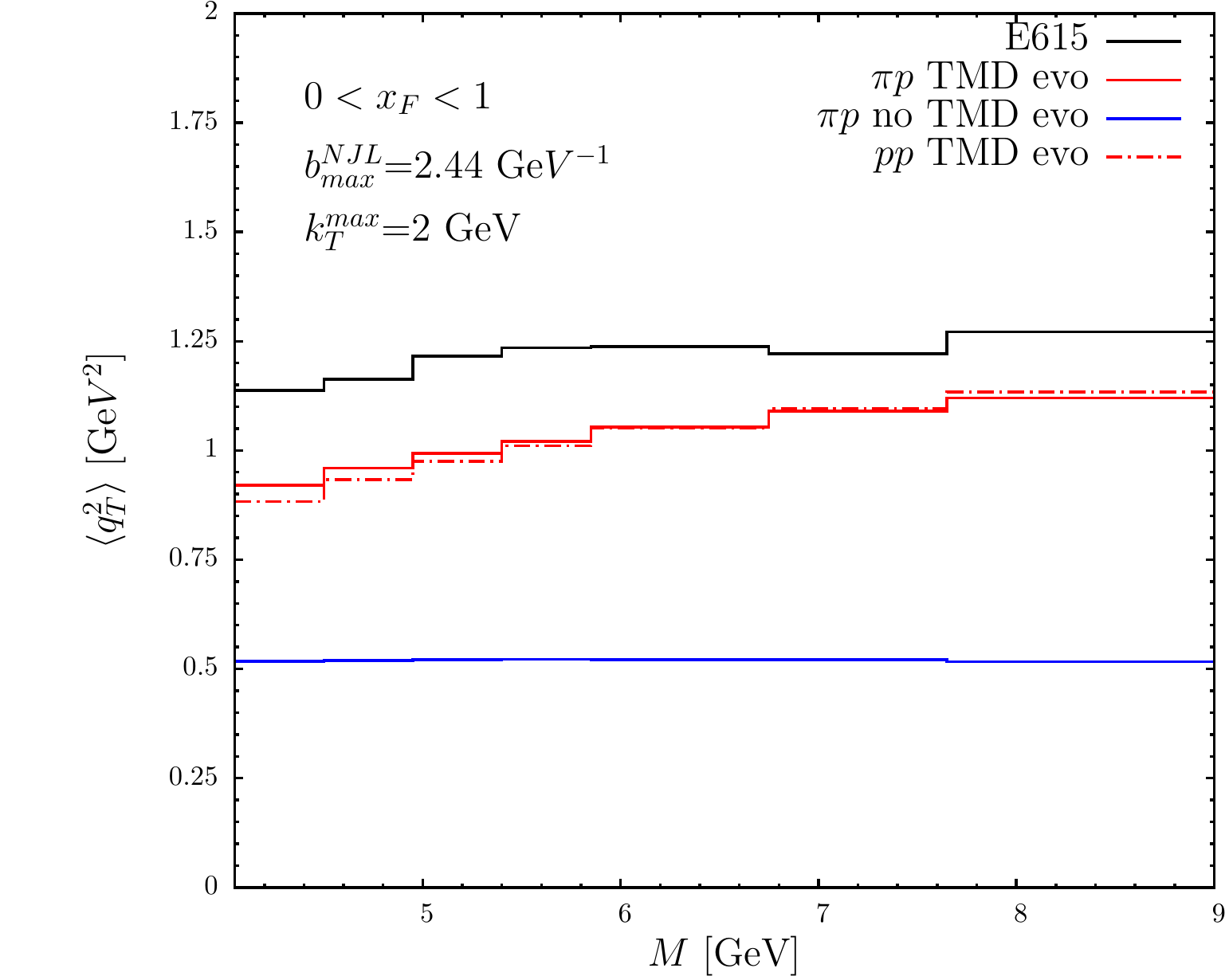}
\caption{Left panel: lepton pair average transverse momentum, $\langle q_T^2 \rangle$, as a function of $x_F$ integrated 
in the mass range $4.0<M/$GeV$ <8.55$. Right panel: $\langle q_T^2 \rangle$ as a function of $M$ integrated 
in the range $0<x_F<1$. Averaged values are obtained integrating both predictions and 
the phenomenological parametrisation of the data up to $q_T^{max}=2$ GeV.}
\label{Fig:6}
\end{center}
\end{figure}

We now turn to the discussion of the perturbative part of the Sudakov form
factor, Eq.~(\ref{sudakov}). The latter, at variance with its non perturbative counter part, 
does not depend upon the type of initial state hadrons involved in the scattering process. 
In principle, the same regularisation procedure should be used both in the Sudakov and in 
the PDFs. This optimum indeed faces some technical problem, for example the call
to PDFs to values outside the boundary of the grid in which they are defined
and the different scales at which the transverse distributions  are assumed to factorise 
on the proton and pion side, respectively.
In order to accomodate all these different settings, we find useful to split the perturbative
form factor in Eq.~(\ref{sudakov}) in a form which allows to use distinct $b_{max}$ on 
the proton and pion side:
\begin{eqnarray}
\label{sudakov_splt}
S_q(Q,b) &\equiv& S_q(Q,b_*,b_{max}^p,b_{max}^\pi)\nonumber\\
&=&\exp 
\left\{ -\frac{1}{2}\int_{\frac{b_0^2}{b^2_*(b_{max}^p)}}^{Q^2} \frac{dq^2}{q^2} 
\left[ A(\alpha_s(q^2)) \;\ln \frac{Q^2}{q^2} + B(\alpha_s(q^2)) \right] \right\}\,  \nonumber\\
&& \times \exp \left\{ -\frac{1}{2}\int_{\frac{b_0^2}{b^2_*(b_{max}^\pi)}}^{Q^2} \frac{dq^2}{q^2} 
\left[ A(\alpha_s(q^2)) \;\ln \frac{Q^2}{q^2} + B(\alpha_s(q^2)) \right] \right\}\,.
\end{eqnarray}
For the proton parameters, we stick to KN05 settings since
the $a_i$'s in Eq. (\ref{fnp}) optimized for $b_{max}^p=1.5$ Ge$\mbox{V}^{-1}$.
On the pion side there is some freedom in adjusting $b_{max}^{NJL}$.
However the pion TMD shows a $x-k_T$ factorised structure only 
at $Q_0^2$, whose numerical value
has been determined in the previous section. 
Therefore we can expect the $b_*$-prescription to involve 
$b_{max}^\pi$ values of the order $b_0/Q_0\sim 2.44$ Ge$\mbox{V}^{-1}$, 
which will be our default value to be used both 
in the Sudakov and in NJL pion parton distributions regularisation. 
\begin{figure}[t]
\begin{center}
\includegraphics[scale=0.6]{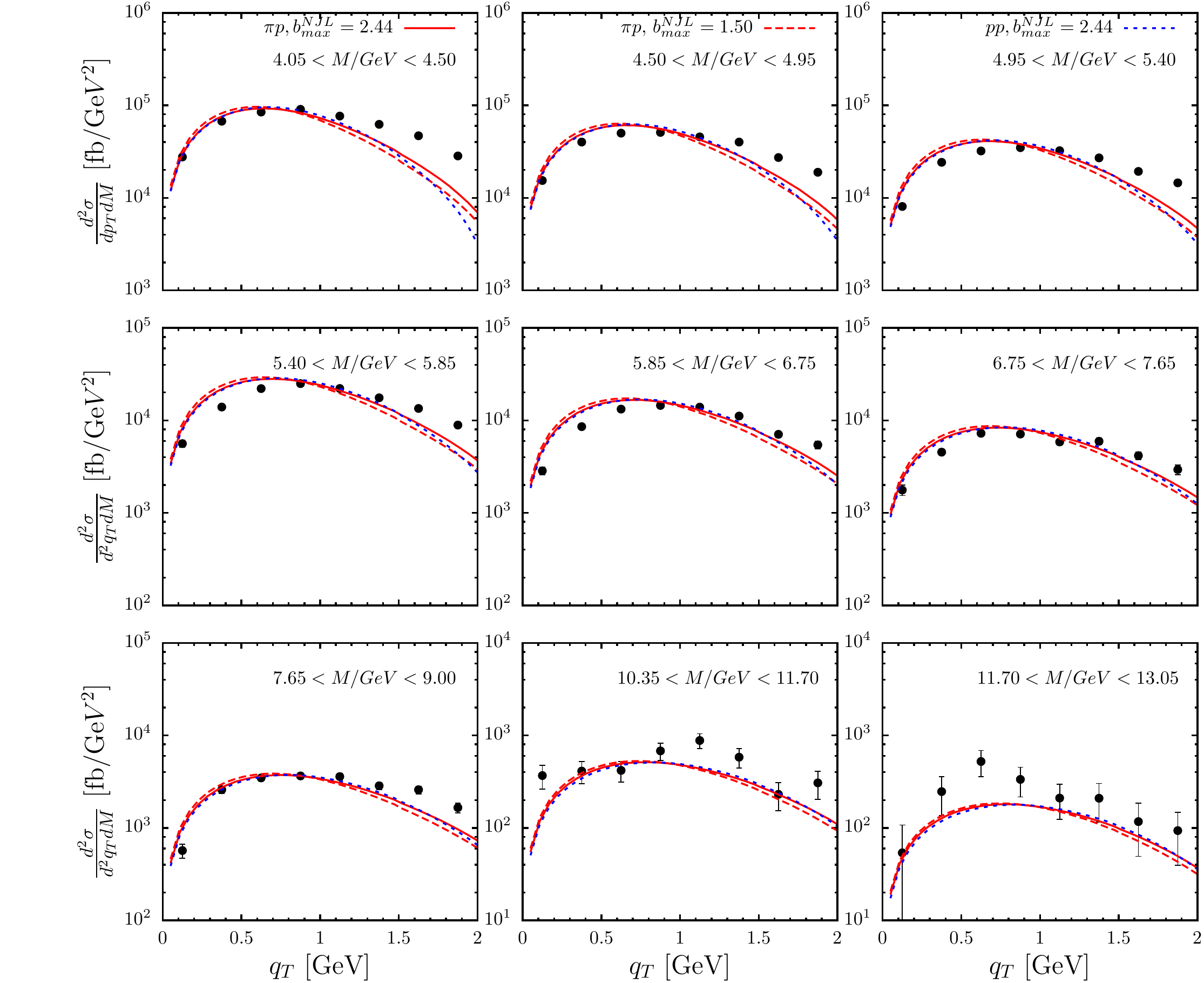}
\caption{Predictions compared to cross sections
in various invariant mass bins of the pair integrated in $0<x_F<1$. 
Data from Refs.~\cite{conway,stirling}}
\label{Fig:4}
\end{center}
\end{figure}
\begin{figure}[t]
\begin{center}
\includegraphics[scale=0.6]{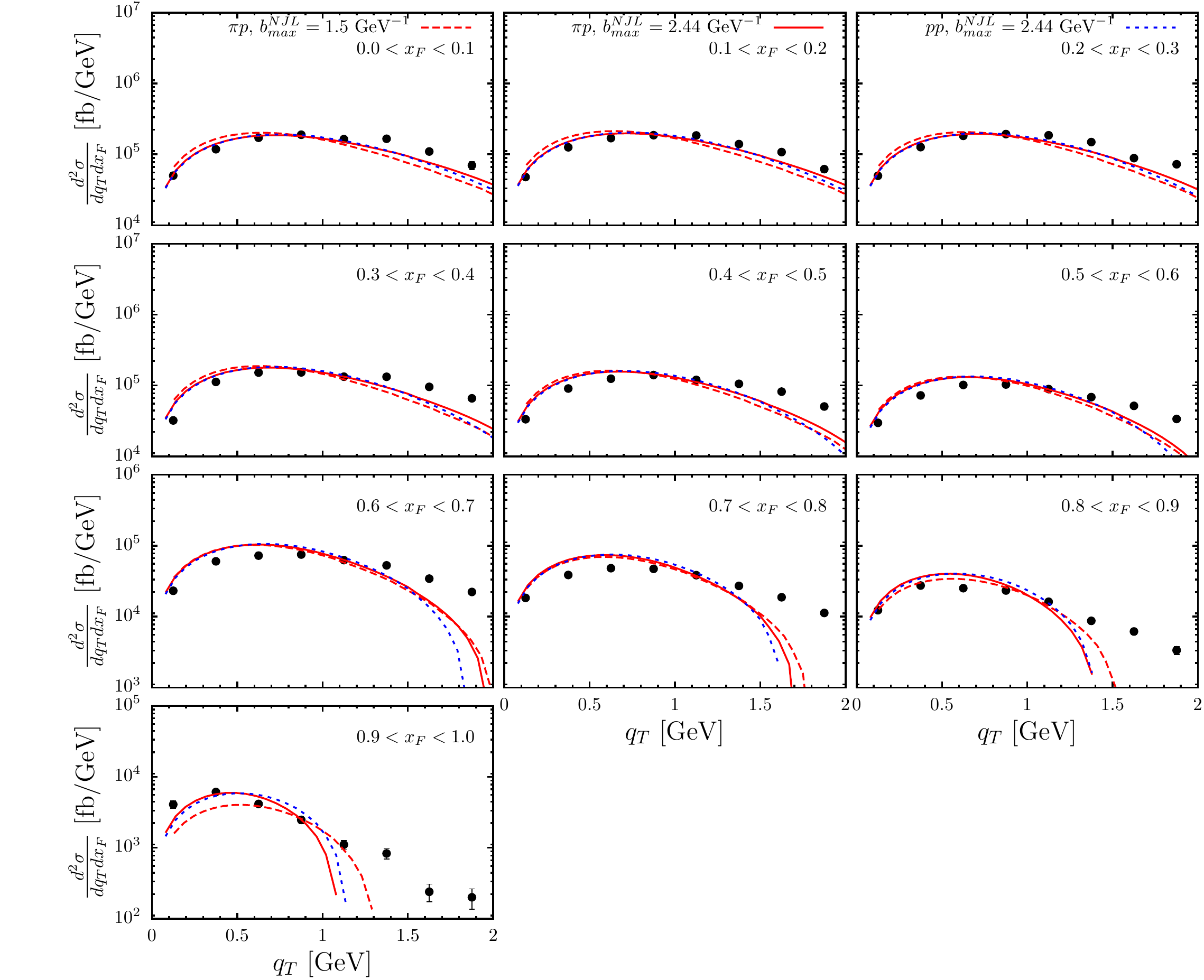}
\caption{Predictions compared to differential cross sections 
in various $x_F$ bins integrated in the mass range $4<M<8.55$ GeV.
Data from Refs.~\cite{conway,stirling}}
\label{Fig:5}
\end{center}
\end{figure}
We now turn to the comparison to lepton pair $q_T$-spectra collected in tables D92-D97
of Refs.~\cite{conway,stirling}, measured in $\pi W$ collisions.
Such data actually refer to less differential
cross sections with respect to the one appearing in Eq.~(\ref{CSS}). In this case 
differential cross sections are integrated over additional variables according to values 
specified in experimental analyses.
We start presenting our results showing, in Fig.~\ref{Fig:4}, cross sections differential 
in $q_T$ integrated in $0<x_F<1$ in various bins of the invariant 
mass of the pair, $M$. 
The comparison is performed up to a $q_T \sim 2$ GeV, where we have checked that the KN05 
gives an adequate description of $pp$ data.
All three different predictions, to be discussed in the following, 
capture the normalisation of the data and share a tendency 
to slightly overestimate the data at very small $q_T$ and to 
underestimate them at larger $q_T$. This effect progressively disappears 
increasing the mass of the lepton pair. Comparing the two curves
corresponding to $b_{max}^{NJL}= 2.44$ Ge$\mbox{V}^{-1}$ and $b_{max}^{NJL}= 1.5$ Ge$\mbox{V}^{-1}$,
one may notice a substantial stability upon variation of the 
regulators on the pion side. On the same plot, in order to investigate 
the sensitivity to the pion transverse distribution, we additionally show 
the predictions obtained by substituting the pion transverse factor, Eq.~(\ref{bsp}), with $\sqrt{S_{NP}^{pp}(b)}$.
As shown in Fig.~\ref{Fig:3}, the non perturbative transverse distributions for the proton and pion differ 
at low scales. The corresponding curve, indicated with $pp$ on the plot, is barely distinguishable from the other two. Such a comparison supports the hypothesis that the effect of the perturbative evolution,
driven by Eq.~(\ref{sudakov_splt}), is to wash away differences  
in the non perturbative structure  found at the hadronic scale. This result implies a reduced sensitivity 
to non perturbative structure.
We proceed our discussion presenting in Fig.~\ref{Fig:5}
the comparison between theory predictions
against the same data, now integrated in the mass range $4<M<8.55$ GeV 
in a number of $x_F$ bins.
We remind the reader that we have verified that the KN05 model gives a satisfactory description of $pp$ data up to $x_F \sim 0.3$. Up to this $x_F$ value, as
shown in the first row of Fig.~\ref{Fig:5}, the description $\pi^-W $ data is fair, as already observed in Fig.~\ref{Fig:4}.
Beyond that range, however,  the width of the theoretical curves decreases more rapidly than 
observed in the data, with data substantially undershooted beyond $q_T \sim 1$ GeV.
This effect is more pronounced as $x_F$ increases.
In this region of relatively large pion fractional momenta it would be tempting to invoke, in order to describe the data, 
an $x$-dependent non perturbative structure. 
Such an interesting hypothesis, however, cannot be tested  unless fixed order contributions at finite $q_T$ are included in the calculation.  
On the other hand we notice that the WLS pion model of Ref.~\cite{Wang:2017zym}, which does not include any additional $x_i$ dependence in the non perturbative form factor, is able to reproduce the data up to $x_F \sim 0.8$. 
On the theoretical side, we would like to mention that, in this range of quite large pion parton 
fractional momenta, the theoretical description of the $q_T$-spectrum can be further improved employing joint resummation techniques 
described in Refs.~\cite{Kulesza:2002rh,Muselli:2017bad}. 
In order to better appreciate how the width of theoretical predictions 
evolves with $x_F$ (and therefore with $x_{\pi}$) and the invariant mass 
of the lepton pair,
we show in Fig.~\ref{Fig:6} the average transverse momentum of the pair,  $\langle q_T^2 \rangle$,  calculated as
\begin{equation}
\langle q_T^2 \rangle=\ddfrac{\int_{x_F^{\mbox{\tiny min}}}^{x_F^{\mbox{\tiny max}}} dx_F \int_{\tau_{\mbox{\tiny min}}}^{\tau_{\mbox{\tiny max}}} d\tau \int_{0}^{q_T^{2,\mbox{\tiny max}}} dq_T^2 \; q_T^2 \; \frac{d^3 \sigma}{dx_F d\tau d q_T^2}}
{\int_{x_F^{\mbox{\tiny min}}}^{x_F^{\mbox{\tiny max}}} dx_F \int_{\tau_{\mbox{\tiny min}}}^{\tau_{\mbox{\tiny max}}} d\tau \int_{0}^{q_T^{2,\mbox{\tiny max}}} dq_T^2  \; \frac{d^3 \sigma}{dx_F d\tau d q_T^2}} \,.
\label{aveqt2}
\end{equation}
Integration limits are provided by experimental conditions.
%
%
For data, indicated by black lines in Fig.~\ref{Fig:6}, 
the phenomenological parametrisation presented in Ref.~\cite{conway} is used.  
For both theory and data, the $\langle q_T^2 \rangle$ is calculated with a maximum value of $q_T^{max}=2$ GeV. 
Theory predictions tend to undershoot the data but, overall, a good 
shape agreement is found. By comparing lines with and without 
TMD evolution (for the latter the perturbative Sudakov $S_q$ is removed from the evaluation of Eq.~(\ref{CSS}))
one can appreciate its large impact on the amount of generated $\langle q_T \rangle $. 
On the same plot, in order to investigate 
the sensitivity to the pion transverse structure, we additionally show 
the predictions obtained by substituting the pion transverse factor, Eq.~(\ref{bsp}), with $\sqrt{S_{NP}^{pp}(b)}$.
As already seen in Fig.~\ref{Fig:4}, difference{\bf s} are minimal, implying
a reduced sensitivity to details of the non perturbative transverse factor.
Therefore if one aims to better appreciate the strictly non perturbative 
form factor, one has to confine in corners 
where TMD evolution is minimised, but still in a perturbative range.
These phase space regions can be identified by extrapolation from the right plot as the one 
at the lowest, but still perturbative, values of the invariant masses of the pair.

\section{Conclusions}

A thorough analysis of DY pair production
in pion-nucleus scattering has been presented.
The  main goal of our work  has been 
the test of model predictions, obtained within the Nambu--Jona-Lasinio
model for the transverse pion structure.
In particular we have focused on the study of differential transverse momentum
spectra of DY pairs produced in 
$pA$ collisions calculated 
in the CSS framework  at NLL accuracy borrowing from the literature 
the longitudinal and transverse proton structure. 
The pion is treated in the Nambu-Jona--Lasinio
model. No further assumption has been made: even the momentum
scale associated to the model is obtained via a minimization procedure of NLO theory to DY experimental longitudinal spectra. The latter turns out to be a low one, in line with that normally used,
which could be predicted within the spirit of the model
without fitting ``a posteriori''.
The agreement found between our pion-nucleus theoretical
cross sections and experimental data is rather successful,
confirming the predictive power
of the NJL model, for both the longitudinal pion parton distributions
and its transverse structure.
We notice that the theory tends to systematically undershoot the data on the higher 
end of the considered $q_T$ interval. All interpretations of this effect, however,
are not conclusive without the inclusion of the finite, fixed order, contributions
which populate the $q_T \sim Q$ region   and are neglected in our calculation.  

The possibility to distinguish between different non perturbative transverse momentum distributions 
in DY data appears instead  more questionable. 
In this complicated scenario, a  possible
strategy  would be the measurement of DY pion-nucleus 
$q_T$-spectra, in bins of $x_F$, at low 
values of the
mass of the pair,
 as the  present study suggests to look into this kinematical
window to emphasize the non-perturbative
content of the pion. Further analyses of the pion non-perturbative form factor, as a function of the hard scale, should be pursued so we could progress on that point.
In the very same window, new data could allow a deeper investigation of the dependence of the non perturbative form factor upon the hard scale of the process.

\section*{Acknowledgments}

We thank P. Nadolsky for a useful mail exchange
about details of the fit presented in Ref. \cite{KN05}.
This work was supported in part by the Mineco
under contract FPA2016-77177-C2-1-P, by GVA-Prometeo/II/2014/066, by
the Centro de Excelencia Severo Ochoa Programme grant SEV-2014-0398 and by UNAM through the PIIF project Perspectivas en F\'isica de Part\'iculas y Astropart\'iculas.
F.A.C. and S.S. thank the Department of Theoretical Physics of the
University of Valencia for warm hospitality and support; F.A.C, A.C.
and S.N. thank the INFN,
sezione di Perugia,  
and the Department of Physics and Geology of the University
of Perugia for warm hospitality and support.

\end{document}